\documentclass[preprint,showpacs,preprintnumbers,amsmath,amssymb,prb]{revtex4}

\usepackage{graphicx}
\usepackage{dcolumn}
\usepackage{bm}

\begin{document}


\title{Cavity dumping of an injection-locked free-electron laser}

\author{Susumu Takahashi}
 \email{susumu@itst.ucsb.edu}
\author{Gerald Ramian}
\author{Mark S. Sherwin}
\affiliation{Department of Physics and Institute for Terahertz Science and Technology, University of California Santa Barbara CA 93106}%

\date{\today}

\begin{abstract}
This letter reports cavity dumping of an
electrostatic-accelerator-driven free-electron laser (FEL) while
it is injection-locked to a frequency-stabilized 240 GHz
solid-state source.  Cavity dumping enhances the FEL output power
by a factor of $\sim$8, and abruptly cuts off the end of the FEL
pulse. The cavity-dumped, injection-locked FEL output is used in a
240 GHz pulsed electron spin resonance (ESR) experiment.
\end{abstract}

                             %
\maketitle

High-power pulsed electron spin resonance (ESR) is an emerging
technique to investigate fast dynamics of biological
molecules~\cite{freed00}. At present, most high-power pulsed ESR
spectrometers operate near the X-band frequency of 9.5 GHz with
kilowatt-level power and $\sim$ 100 nanosecond time resolution. A
trend in the evolution of next generation pulsed ESR is for higher
magnetic fields and frequencies for both finer spectral and time
resolution. Because the linewidth of ESRs tends to be extremely
narrow, the source radiation must have a correspondingly narrow
bandwidth and stable frequency. High-power pulsed ESR, using
few-ns pulses to rapidly manipulate spins for spin-echo and
related experiments, has been demonstrated at 95 GHz at Cornell
using a kW-power klystron-based source~\cite{hofbauer04}. Another
klystron based-pulsed ESR system has been built at St. Andrews,
UK~\cite{standrews}.

A bottleneck for a higher-frequency pulsed ESR spectrometer is a
scarcity of sources with high power and narrow bandwidth.
Commercially-available klystron-amplifiers operating between 200
and 300 GHz can currently generate several tens of watts of peak
power. Gyrotron sources with \textgreater 100W peak power below
200 GHz have been developed for dynamic nuclear polarization (DNP)
and ESR experiments~\cite{hall97}. Gyrotrons generating several
tens of watts of continuous wave (cw) output at 1 THz have also
been made~\cite{fukuiTHz}. However, the frequency of a gyrotron is
usually not tunable. There are also commercially-available
solid-state based sources which generate 10s of mW near 240 GHz,
and operate at higher frequencies with lower power. The
solid-state sources typically consist of a microwave frequency
synthesizer, microwave amplifiers and frequency multipliers.
Free-electron lasers (FELs) are also good candidates as sources of
tunable high-power pulsed radiation at terahertz and sub-terahertz
frequencies. The UC Santa Barbara (UCSB) FELs~\cite{ucsbfel},
driven by an electrostatic accelerator, produce several
micro-second pulses of quasi-cw radiation whose frequency is
tunable from 120 GHz to 4.7 THz. Lasing on a single-mode with
sub-megahertz line-width has recently been demonstrated by using
the recently-installed injection-locking
system~\cite{takahashi07}. The UCSB millimeter-wave FEL (MM-FEL)
typically generates hundreds of watts at 240 GHz, which is being
used to develop a FEL-driven pulsed ESR
spectrometer.~\footnote{This relatively low power (for a FEL)
arises because the UCSB MM-FEL was optimized to run at
significantly higher frequencies, where powers in excess of 10 kW
are possible without cavity dumping.}

In this article, we demonstrate the operation of a cavity dump
coupler (CDC) in combination with injection-locking of the UCSB
MM-FEL. This is a key achievement for FEL-powered pulsed ESR. The
CDC does two important things. First it increases peak power to to
above 1 KW levels, and second it abruptly switches that power off.
The latter is critical in improving the signal-to-noise ratio of
the pulsed ESR spectrometer.

\begin{figure}
\includegraphics[width=80 mm]{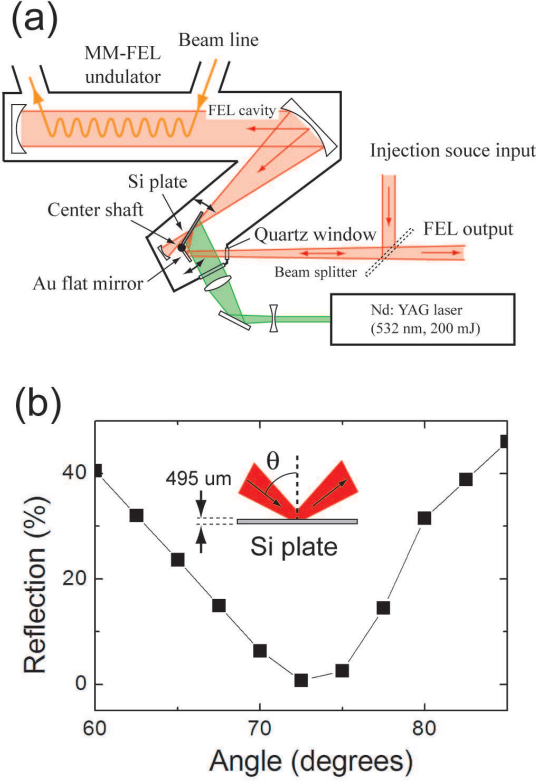}
\caption{\label{fig:CDCsetup} (a)Schematic of the injection
locking system for the UCSB MM-FEL. The injection source, the
isolator and a tunable 100 MHz synthesizer are located in
free-electron laser lab. The FEL outputs are sent to the users lab
through vacuum optical transport system. (b)Reflectivity
characteristics of optimized Si plate with 495 $\mu$m thickness.}
\end{figure}
Fig.~1(a) shows a schematic of the UCSB MM-FEL with the CDC. The
MM-FEL which covers from 120 to 890 GHz employs a 6.25 m-long
folded cavity, a Si plate coupler, and optical and THz access in a
room-temperature high-vacuum housing. The coupler consists of a
high-resistivity silicon (Si) plate. Because the band gap of Si is
much greater than photon energy at THz frequencies, the Si plate
is normally transparent to THz radiation. However, when the Si
wafer is illuminated by a high flux of radiation with photon
energy greater than the band gap, then the plasma frequency of
excited carriers exceeds the THz frequency, and the Si plate
becomes highly reflective. The excitation of carriers on Si
surface is so fast that the actual switching time of the Si
coupler is determined by the pulse characteristics of the
excitation laser. The switch remains "on", for several
microseconds, limited by the recombination time of the
photo-excited carriers. This photo-activated semiconductor
switching technique was first demonstrated with polycrystalline
germanium and later silicon~\cite{alcock75, salzmann83}. In
addition, the technique is regularly used for a fast pulse slicer
switch of THz waves in the UCSB users lab~\cite{hegmann00,
doty04}.

As shown in Fig.~1(a), the MM-FEL CDC system has a capability to
change the angle of the Si plate with respect to the incident FEL
radiation away from Brewster's angle. This provides variable
coupling to maximize the FEL output power in normal operation
({\it i.e.} without cavity dumping). The thickness of the Si plate
was carefully chosen as 495 um in order to provide adequate
coupling as a function of angle as shown in Fig.~1(b). In use, the
Si angle is remotely controlled by a stepper motor which rotates a
center shaft connected to the Si plate and flat mirror to maintain
constant angle and position of the output beam. In the FEL output
port, we employ a quartz window whose thickness is optimized at
240 GHz for minimal reflection.

We employ a frequency-doubled Nd:YAG laser (Big Sky Laser
CFR200~\cite{bigsky}) to switch the Si coupler. The laser emits
pulses with 532nm of wavelength, 7$\pm$2 ns full-width at
half-maximum (FWHM) and $\sim$ 200 mJ of energy. Timing of the
switching is controlled within $\pm$ 500 ps accuracy by precisely
triggering the Q-switch and flash lamp of the Nd:YAG laser with
FEL advanced trigger pulses using a Stanford research systems
delay generator DG645. The 532 nm green radiation illuminates the
Si plate through an anti-reflection coated window. The CDC works
in conjunction with the recently installed injection-locking
system~\cite{takahashi07} for single mode operation. The
injection-source is coupled to the FEL cavity through the Si
coupler. Injection-locking was achieved even with the minimum Si
coupling at Brewster's angle.

A CDC has been previously demonstrated in the UCSB
FEL~\cite{kaminski90}. Although the experiment clearly showed the
switching of the FEL with a Si plate, the performance of the CDC
was much lower than what is anticipated because of inadequate
drive laser energy. The current CDC system has been completely
upgraded with a proper drive laser and Si wafer. In addition, a
combination with injection locking makes the FEL a suitable source
for high-power pulsed ESR experiments.

\begin{figure}
\includegraphics[width=70 mm]{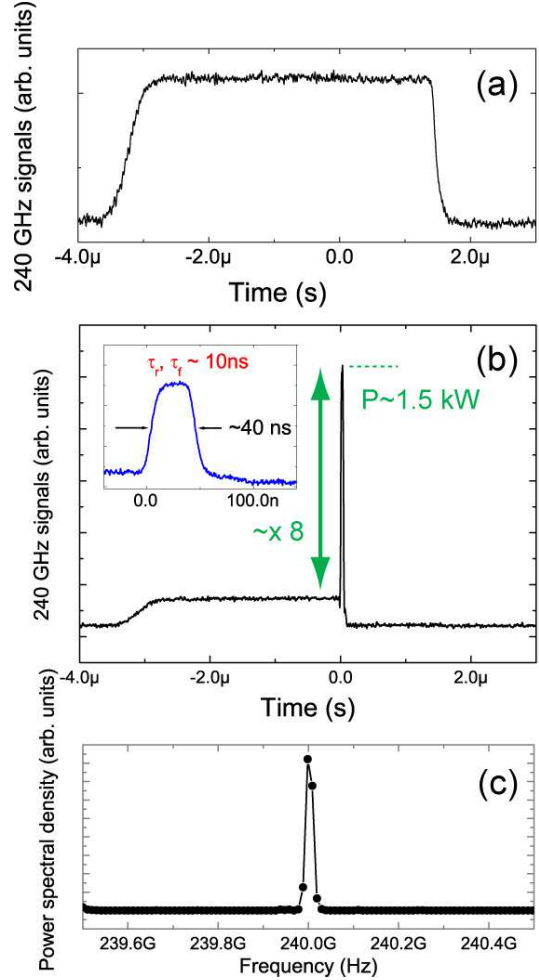}
\caption{\label{fig:CDCtrace} (a) 240 GHz FEL output without the
CDC operation as a function of time. The signal was measured using
a video-mode detector from VDI. The FEL signal was attenuated down
to tens of microwatts in order to protect and not to saturate the
detector. Digitization noise was filtered without affecting rise
and fall times using a 6th order Savitsky$-$Golay smoothing
filter. (b) 240 GHz FEL output with CDC operation. The inset shows
a detailed shape of the CDC FEL signal. Rise and fall time
($\tau_r$ and $\tau_f$) of the CDC FEL $\tau\sim$ 10 $ns$ is given
by a time difference between 90 \% and 10 \% of the signal. (c)
Power spectral density of a CDC FEL pulse with the
injection-locking technique. The width of the CDC FEL spectrum is
approximately 22 MHz.}
\end{figure}
Fig.~2(a) and (b) show the recorded 240 GHz FEL output signals
without and with CDC respectively. A $\sim$4.7 $\mu$s pulse, in
the normal FEL mode, was observed in the FEL user lab as shown in
Fig.~2(a). The signal shows no evidence of mode-beating effects
because of the operation of the injection-locking system. The peak
power of the FEL pulse, measured using a Thomas-Keating energy
meter~\cite{tk}, was $\sim$ 190 W in the user lab. When the CDC is
operated, the FEL output trace is drastically different, as shown
in Fig.~2(b). The FEL radiation rises at the time$=-3$ $\mu$s
point, and the CDC drive laser is triggered later at time$=$0. In
this case, the measured peak power of the CDC FEL is 1.5 kW which
is approximately 8 times more than the normal FEL power. As shown
in the inset of Fig.~2(b), the radiation lasts $\sim$40 ns which
corresponds to the L=6.5m length of the FEL cavity ($t=c/2L$,
where $c$ = speed of light). The rise and fall times of the CDC
FEL pulse corresponds to the switching time of the Si plate. In
the present case, we found the duration of the both rising and
falling edge to be $\sim$10 ns, which is similar to the FWHM of
the excitation laser pulse. As explained later, this short decay
time is very important to reduce background noise after the pulse.
The spectrum of the CDC FEL radiation was measured by a heterodyne
spectrometer~\cite{takahashi07}. As shown in Fig.~2(c), the CDC
pulse spectrum remains single mode as enforced by the
injection-locking system. The observed spectral line-width is
$\sim$ 22 MHz, consistent with the Fourier transform limit.

\begin{figure}
\includegraphics[width=90 mm]{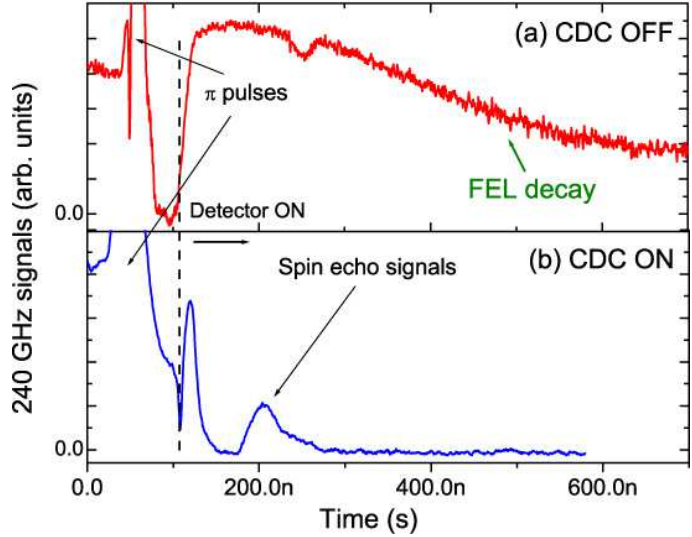}
\caption{\label{fig:SpinEcho} Spin echo signals by Hahn echo
sequence ($\pi$/2-$\tau$-$\pi$-$\tau$-echo) with the normal FEL
(a) and CDC FEL (b). Data was taken at 1.5 Kelvin and the sample
is Fe8 single molecule magnets. Details of this experiment will be
presented elsewhere.}
\end{figure}
A FEL-powered pulsed ESR spectrometer is currently being developed
with the injection-locked UCSB FEL. The system includes a new
pulse slicer optimized at 240 GHz, a home-built detector, and
protection for the detection system. Details of this setup will be
presented elsewhere. As a novel application of the CDC FEL in
pulsed ESR, Fig.~3 shows the first spin echo signal taken by the
FEL-powered pulsed ESR spectrometer. In the case of FEL-powered
pulsed ESR, the frequency of the applied pulses and detected
signals is the same, therefore isolation of the applied pulses is
very critical. However, because the detected ESR signals are often
less than nano-watts, while the excitation pulses are hundreds of
watts, the isolation of the signal from the excitation pulse is
extremely difficult. Fig.~3(a) shows a spin echo taken without
cavity dumping normal FEL. Two Excitation pulses ($\pi/2$ and
$\pi$) were applied before time=50 ns and the detection system was
switched ON at 100 ns. After time=100 ns, as shown in Fig.~3(a),
spin echo signals are hidden and distorted by a large background
and noise due to imperfect pulse slicing in the pulse slicer. This
issue can be solved using the CDC FEL. Because the CDC shuts off
the FEL radiation quickly and completely, background and noise
after the CDC FEL pulse are extremely small. Fig.~3(b) shows spin
echo measurement with the CDC FEL, where timing of the spin echo
is set at 200 ns. As shown in the figure, background noise is
negligible and the CDC FEL gives a clear and non-distorted spin
echo signal.

In summary, we have demonstrated cavity dumping of an injection
locked-FEL. The CDC FEL radiation shows close to an order of
magnitude enhancement in the FEL output power. The CDC FEL also
gives abrupt falling edge of the FEL radiation which is critical
for background-free pulsed ESR measurement. The combination of
cavity dumping and injection locking in FELs enables one to use
FELs for pulsed ESR as well as nonlinear spectroscopy of other
systems with extremely narrow linewidths, like dilute molecular
gases and Rydberg atoms.

We thank David Enyeart, Devin Edwards and Louis-Claude Brunel for
support of the FEL-powered ESR operation. This work was supported
by research grants from NSF (DMR-0520481, DMR-0703925 and
CHE-0821589) and the W. M. Keck Foundation.


%
%
%
%
%
%
%

\end{document}